\documentclass[12pt]{article}
\usepackage{amssymb,amsmath,epsfig}

\begin{document}

\title{\bf Collapse and Expansion of Anisotropic Plane Symmetric Source}

\author{G. Abbas \thanks{ghulamabbas@ciitsahiwal.edu.pk}
\\Department of Mathematics, COMSATS Institute\\
of Information Technology, Sahiwal-57000, Pakistan.}

\date{}
\maketitle
\begin{abstract}
This paper deals with the collapse and expansion of relativistic
anisotropic self-gravitating source. The field equations for
non-radiating and non-static plane symmetric anisotropic source have
been evaluated. The non-radiating property of the fluid leads to
evaluation of the metric functions. We have classified the dynamical
behavior of gravitational source as expansion and collapse. The
collapse in this case leads to the final stage without the formation
of apparent horizons while such horizons exists in case of spherical
anisotropic source. The matching of interior and exterior regions
provides the continuity of masses over the boundary surface.
\end{abstract}
{\bf Keywords:} Anisotropic Fluids; Expansion; Gravitational
Collapse.\\
 {\bf PACS:} 04.20.Cv; 04.20.Dw
\section{Introduction}

Oppenheimer and Snyder (1939) are the pioneers who studied
gravitational collapse of dust. This was the special problem because
dust was unrealistic matter and one cannot ignore the effects of
pressure on singularity formation during gravitational collapse. A
more analytic analysis was made by Misner and Sharp (1964) with
perfect fluid in the interior region of a star. They formulated the
dynamical equations governing adiabatic relativistic collapse. In
both cases, vacuum was taken in the exterior region of a star. Also,
Misner and Sharp (1965) formulated the problem of collapse for
anisotropic fluid. Since then there has been a growing interest to
study the gravitational collapse of anisotropic fluid spheres (
Herrera et al. 2008a and Herrera et al. 2008 b). The general
solution of anisotropic models has attained a considerable interest
in Einstein theory of gravity because of applications to stellar
collapse of spheres ( Bowers et al. 1974, Cosenza et al. 1981, Bayin
1982, Sharif and Abbas 2013 and Bondi 1992). Barcelo et al. (2008)
have studied the gravitational collapse in semiclassical theory of
gravity. They pointed out that in this approach Hawking radiations
might prevent the formation of trapped surfaces and apparent
horizons. In this way they proposed a new class of collapsing
objects with no horizons. Several authors (Nojiri and Odintov 2005,
 Nojiri 2006, Cognola et al. 2007 and Gasperini et al. 1993) have discussed the anisotropy of
 dark energy in modified theories
of gravity. Also, Herrera and Santos (1997) have investigated the
some properties of anisotropic self-gravitating system and
determined the stability of the perturbed system. Herrera et al.
(2008a) have discussed the possibility of a single generated
function, that possibility has been developed in this paper.

The method of generating the gravitational collapse of non-adiabatic
collapse was initially developed by the Glass (1981). In this method
he used a perfect fluid static or non-static solution of shear-free
collapse to a shear-free collapse model with radial heat flow. In a
recent paper, the same author (Glass 2013) has constructed a
solution of adiabatic anisotropic sphere which leads to either
expansion and collapse depending on he choice of initial data. In
the present paper, we extend this work to plane symmetric objects.
The models exhibiting plane symmetry can be used as test-bed for
numerical relativity, quantum gravity and contribute for examining
CCH and hoop conjecture among other important issues.

The significant features of gravitational collapse motivated us to
extend the work of Glass (2013) to plane symmetry. The models
exhibiting plane symmetry can be used as test-bed for numerical
relativity, quantum gravity and contribute for examining CCH and
hoop conjecture among other important issues. Sharif and his
collaborators (Sharif and Zaeem 2012 and Sharif and Yousaf 2012)
argued that plane symmetric models are more suitable than spherical
for discussing early stage of evolution of the universe. In this
paper, we present a systematic pattern of Glass (2013) to analyze
plane symmetric solutions which exhibit the expansion or collapse.
We have determined the generating solutions, which in case of
collapse imply the absence of trapped surfaces and apparent
horizons. The matching conditions provide the continuity of
gravitational masses in the interior and exterior regions of a
gravitating object.

This paper is organized as follows: In section \textbf{2}, we
present the anisotropic source and Einstein field equations. Section
\textbf{3} deals with the generating generating solutions which
represents collapse as well as expansion of the plane symmetric
spacetime. The matching of interior anisotropic fluid has been
performed with the exterior vacuum solution in section \textbf{4}.
In the last section, we summaries the results of the paper.

\section{Anisotropic Gravitating Source and Field Equations}

We consider a plane symmetric distribution of the fluid, bounded by
a hypersurface $\Sigma$. The line element for the interior region
has the following form
\begin{equation}\label{1}
ds^{2}_{-}=-A^2(t,z)dt^{2}+B^2(t,z)(dx^{2}+dy^{2})+C^2(t,z)dz^{2},
\end{equation}
where we have assumed co-moving coordinates inside $\Sigma$. It is
assumed that fluid is locally anisotropic. The energy-momentum
tensor for such a fluid is given by
\begin{equation}\label{2}
T_{ab}=(\mu+P_{\perp})V_{a}V_{b}+P_{\perp}g_{ab}+(P_{z}-P_{\perp})\chi_{a}\chi_{b},
\end{equation}
here $P_{z}$ is the pressure in $z$-direction, $P_{\perp}$ is the
pressure perpendicular to $z$-direction (i.e., $x$ or $y$ direction)
and $\chi^{a}$ is a unit vector in $z$-direction. Moreover, these
quantities satisfy the relations
\begin{equation}\label{3}
V^{a}V_{a}=-1,\quad \chi^{a}\chi_{a}=1,\quad \chi^{a}V_{a}=0.
\end{equation}
For Eq.(1) to be co-moving, we can have
\begin{equation}\label{6}
V^{a}=A^{-1}\delta^{a}_{0},\quad \chi^{a}=C^{-1}\delta^{a}_{3}.
\end{equation}

The expansion scalar is
\begin{equation}\label{7}
 \Theta=\frac{1}{A}(\frac{2\dot{B}}{B}+\frac{\dot{C}}{C}),
\end{equation}
where dot and prime denote differentiation with respect to $t$ and
$z$ respectively. We define a dimensionless measure of anisotropy as
\begin{equation}\label{9}
\Delta a=\frac{P_z-P_{\perp}}{P_z}.
\end{equation}
The Einstein field equations yield the following set of equations
\begin{eqnarray}\label{10}
8\pi\mu A^{2}
&&=\frac{\dot{B}}{B}\left(\frac{2\dot{C}}{C}+\frac{\dot{B}}{B}\right)
+\left(\frac{A}{C}\right)^{2}\left(\frac{-2B''}{B}
+\left(\frac{2C'}{C}-\frac{B'}{B}\right)\frac{B'}{B}\right),
\\\label{11}
&&-\frac{\dot{B}'}{B}+\frac{A'\dot{B}}{AB}+\frac{\dot{C}B'}{CB}=0,
\\\label{12}
8\pi P_{\perp}B^2&&=-\left(\frac{B}{A}\right)^{2}
\left[\frac{\ddot{B}}{B}+\frac{\ddot{C}}{C}-\frac{\dot{A}}{A}\left(\frac{\dot{B}}{B}+\frac{\dot{C}}{C}\right)
+\frac{\dot{B}\dot{C}}{BC}\right]\nonumber\\
&&+\frac{B^{2}}{C^{2}}\left[\frac{A''}{A}+\frac{B''}{B}-\frac{A'}{A}\left(\frac{C'}{C}
-\frac{B'}{B}\right)-\frac{B'C'}{BC}\right],
\\\label{13}
8\pi P_{z}C^2&&=-\left(\frac{C}{A}\right)^{2}
\left[\frac{2\ddot{B}}{B}+\left(\frac{\dot{B}}{B}\right)^{2}-\frac{2\dot{A}\dot{B}}{AB}\right]\nonumber\\
&&+\left(\frac{B'}{B}\right)^{2}+\frac{2A'B'}{AB}.
\end{eqnarray}

Now by inspection, we find that for arbitrary functions $h(t)$ and
$f(z)$, Eq.(\ref{11}) can be satisfied identically if $A$ and $C$
have following form
\begin{equation}\label{14}
A=h(t)\frac{B}{B^{\alpha}}\quad C=f(z){B^{\alpha}}.
\end{equation}
In this case interior metric (1) can be written as
\begin{equation}\label{15}
ds^{2}_{-}=-\left(\frac{B}{B^{\alpha}}\right)^2dt^{2}+B^2(t,z)(dx^{2}+dy^{2})+{B^{2\alpha}}(t,z)dz^{2},
\end{equation}
where $h(t)$ and $f(z)$ have been absorbed in $dt$ and $dz$,
respectively.

Now using Eq.(\ref{12}), we have following form of expansion scalar
\begin{equation}\label{16}
\Theta=(2+\alpha)B^{(1-\alpha)}.
\end{equation}
For $\alpha>-2$ and $\alpha<-2$, we have expanding and collapsing
regions.

Using Eq.(\ref{12}), we have the following form of matter components

\begin{eqnarray}\label{17}
8\pi\mu
&=&(1+2\alpha)B^{2{\alpha}-2}-\frac{1}{B^{2\alpha}}\left[\frac{2B''}{B}+(1-2\alpha)\left(\frac{B'}{B}\right)^2\right]+\frac{1}{B^2},
\\\label{18}
8\pi
P_{z}&=&(1+2\alpha)B^{2{\alpha}-2}+\frac{1}{B^2}+\frac{1}{B^{2\alpha}}\left[(2\alpha-1)\left(\frac{B'}{B}
\right)^2-\frac{2\dot{B}'}{\dot{B}}\frac{B'}{B}\right],
\\\nonumber
8\pi P_{\perp}&=&-\alpha(1+2\alpha)B^{2{\alpha}-2}\\\nonumber
&+&\frac{1}{B^{2\alpha}}\left[(1-\alpha)\frac{B''}{B}-(3\alpha-1)\frac{\dot{B}'}{\dot{B}}\left(\frac{B'}{B}
\right)+\frac{\dot{B}''}{B}+\alpha(2\alpha-1)\left(\frac{B'}{B}
\right)^2\right].\\\label{19}
\end{eqnarray}
For particular values of $B(z,t)$ and $\alpha$, one can find an
anisotropic configuration. The sectional curvature mass which is
known as Taub's mass function (Zannias 1990) for plane symmetric
spacetime can be written as
\begin{equation*}
m(t,z)=\frac{(g_{11})^{3/2}}{2}R^{12}_{12}.
\end{equation*}
Replacing the values of $g_{11}$ and $R^{12}_{12}$ along with
$A=\frac{B}{B^{\alpha}}\quad C={B^{\alpha}}$ the mass function
becomes
\begin{equation}\label{c1}
\frac{2m(t,z)}{B}=\left({B^{2{\alpha}}}-\frac{B'^{2}}{{B^{2\alpha}}}\right).
\end{equation}
Following (Glass 2013), we can find two rapping scalars which are
given by
\begin{equation}\label{c2}
\kappa_{1}=\frac{{B^{{\alpha}}}}{B}+\frac{B'}{B^{\alpha+1}}, \quad
\kappa_{2}=\frac{{B^{{\alpha}}}}{B}-\frac{B'}{B^{\alpha+1}}.
\end{equation}
At ${B'}=B^{2\alpha}$, we get $m(z,t)=\kappa_{2}=\kappa_{2}=0$. This
implies the absence of apparent horizons, i.e., gravitational
collapse does not lead to trapping situation. The no trapping
condition ${B'}=B^{2\alpha}$, has the integral
\begin{equation}\label{c3}
B^{(1-2\alpha)}_{no~trap}=(1-2\alpha)z+H(t),
\end{equation}
where $H(t)$ is arbitrary function. When he trapped condition
${B'}=B^{2\alpha}$ is applied into Eqs.(\ref{17}), (\ref{18}) and
(\ref{19}), we get
\begin{equation}\label{c4}
8\pi \mu=B^{-2}_{no~trap},\quad 8\pi
P_{z}=B^{-2}_{no~trap},\quad8\pi P_{\perp}=0.
\end{equation}

\section{Generating Solution}

For negative and positive values of $\alpha$, we have collapsing and
expanding solutions as follows:

\subsection{Collapse with $\alpha=-\frac{5}{2}$}
For collapse, the rate of expansion must be negative, from
Eq.(\ref{7}), $\Theta<0$, when $\alpha<-2$, we assume that
$\alpha=-\frac{5}{2}$ and the condition ${B'}=B^{2\alpha}$, leads to
${B'}=B^{-5}$, the integral of this equation is
\begin{equation}\label{c5}
B_{no~trap}=(6z+h_1(t))^{\frac{1}{6}},
\end{equation}
where $h_1(t)$ is arbitrary function of $t$. For
$\alpha=-\frac{5}{2}$, Eqs.(\ref{10}), (\ref{11}) and (\ref{12})
give
\begin{eqnarray}\label{c6}
8\pi\mu
&=&-4B^{-7}-2{B^{5}}\left[\frac{2B''}{B}+3\left(\frac{B'}{B}\right)^2\right]+\frac{1}{B^2},
\\\label{c8}
8\pi P_{z}&=&-4B^{-7}+\frac{1}{B^2}-2B^{5}\left[3\left(\frac{B'}{B}
\right)^2-\frac{2\dot{B}'}{\dot{B}}\frac{B'}{B}\right],
\\\label{c9}
8\pi
P_{\perp}&=&-10B^{-7}+{B^{5}}\left[\frac{7}{2}\frac{B''}{B}+\frac{17}{2}\frac{\dot{B}'}{\dot{B}}\left(\frac{B'}{B}
\right)+\frac{\dot{B}''}{B}+15\left(\frac{B'}{B} \right)^2\right].
\end{eqnarray}
For the different solution, we assume
$B_{no~trap}=k(6z+h_1(t))^{\frac{1}{6}},$ the above equations in
this case reduces to
\begin{eqnarray}\label{c10}
8\pi\mu &=&4k^5(1-k^{-12})(6z+h_1)^{-7/6}+k^{-2}(6z+h_1)^{-1/3},
\\\label{c11}
8\pi P_{z}&=&4k^5(1-k^{-12})(6z+h_1)^{-7/6}+k^{-2}(6z+h_1)^{-1/3},
\\\label{c12}
8\pi P_{\perp}&=&10(k^4-1)(6z+h_1)^{-7/6}.
\end{eqnarray}
When $k=1$, above values reduce to the values in Eq.(\ref{c1}),
therefore we assume $k\neq1$ in our discussion. The dimensionless
measure of anisotropy defined by Eq.(\ref{9}) is
\begin{equation}
\triangle a=1+\frac{10k^2(1-k^4)}{4k^7(1-k^{-12})+(6r+h_1)^{5/6}}.
\end{equation}
\begin{figure}
\center\epsfig{file=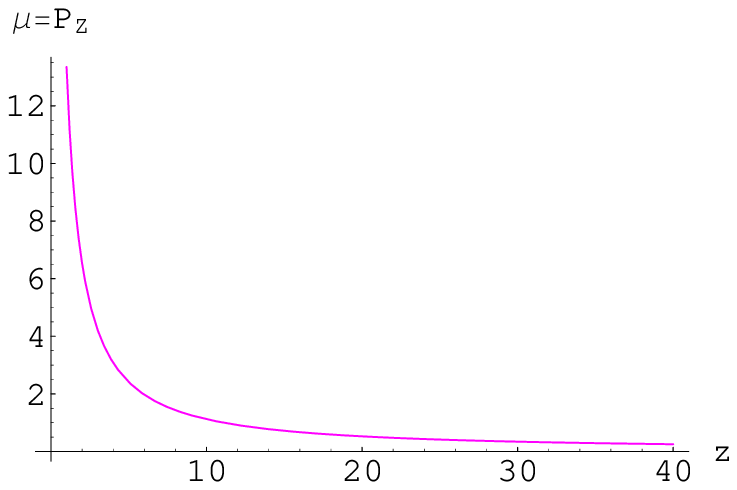, width=0.45\linewidth} \epsfig{file=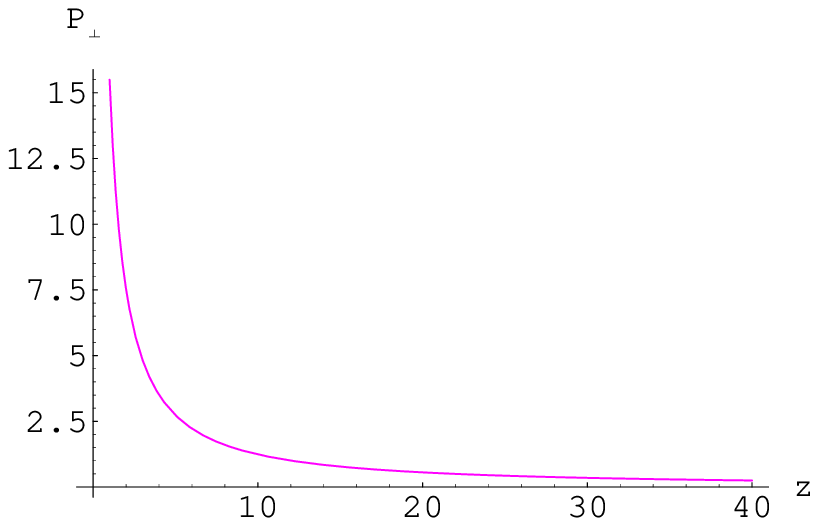,
width=0.45\linewidth}\caption{Both graphs have been plotted for
$k=2, h_1=1$.}
\end{figure}
\begin{figure}
\center\epsfig{file=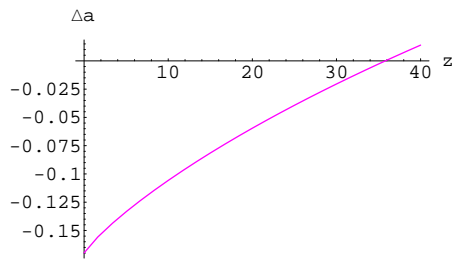, width=0.45\linewidth} \caption{This graph
shows that anisotropy increases from negative value to some positive
value for $k=2, h_1=1$.}
\end{figure}

\subsection{Expansion with $\alpha=\frac{3}{2}$}

For expansion, the rate of expansion must be positive, from
Eq.(\ref{7}), $\Theta>0$, when $\alpha\geq0$, also we assume that
\begin{equation}
B=(z^2+{z_0}^2)^{-1}+h_2(t),
\end{equation}
where $h_2(t)$ and $z_0$ are arbitrary function and constant
respectively.

For $\alpha=\frac{3}{2}$, we get

\begin{eqnarray}\label{c6}
8\pi\mu
&=&4B-2{B^{-3}}\left[\frac{B''}{B}-\left(\frac{B'}{B}\right)^2\right]+\frac{1}{B^2},
\\\label{c8}
8\pi P_{z}&=&4B+B^{-2}+2B^{-3}\left[\left(\frac{B'}{B}
\right)^2-\frac{\dot{B}'}{\dot{B}}\frac{B'}{B}\right],
\\\label{c9}
8\pi
P_{\perp}&=&-6B+{B^{-3}}\left[-\frac{1}{2}\frac{B''}{B}-\frac{7}{2}\frac{\dot{B}'}{\dot{B}}\left(\frac{B'}{B}
\right)+\frac{\dot{B}''}{B}+3\left(\frac{B'}{B} \right)^2\right].
\end{eqnarray}
With $F(t,z)=1+h_2(t)(z^2+{z_0}^2)$ and $B=\frac{F}{(z^2+{z_0}^2)}$,
the density and pressures in this case are
\begin{figure}
\center\epsfig{file=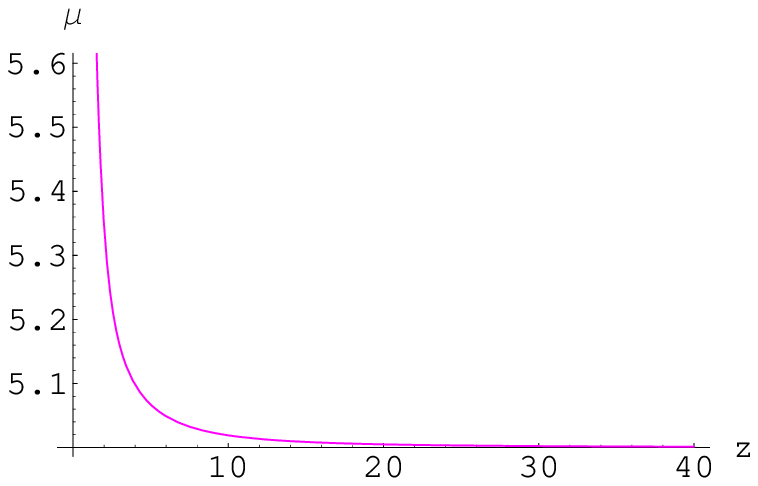, width=0.45\linewidth} \epsfig{file=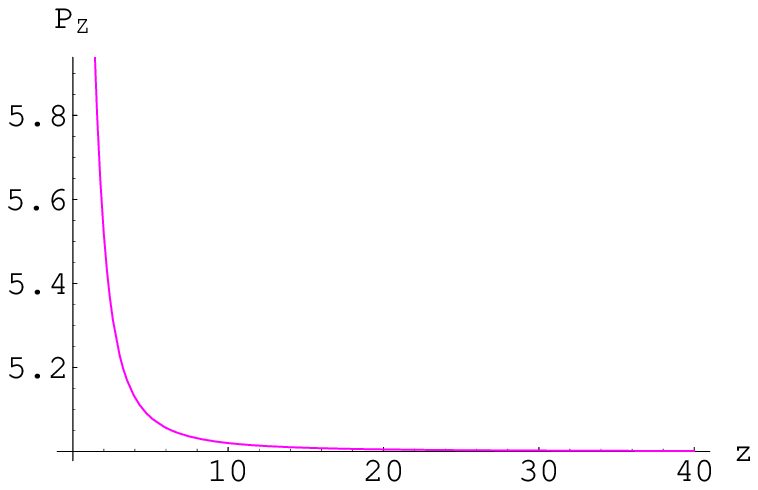,
width=0.45\linewidth}\caption{This graph is plotted for $h_2=1,
z_0=1$.}
\end{figure}
\begin{figure}
\center\epsfig{file=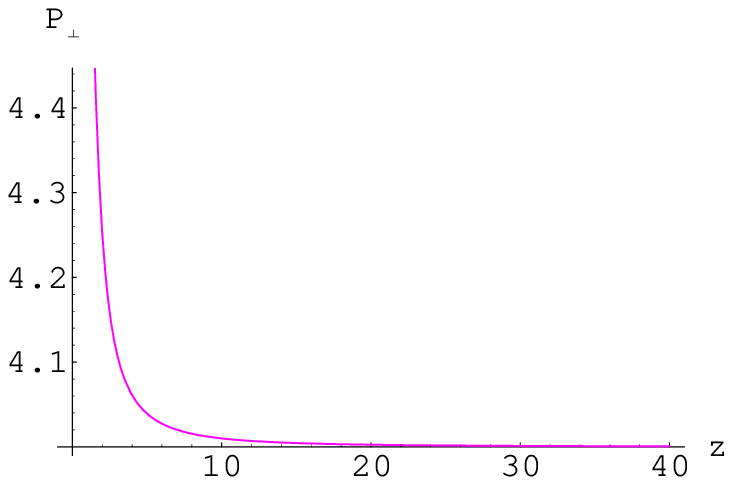, width=0.45\linewidth} \epsfig{file=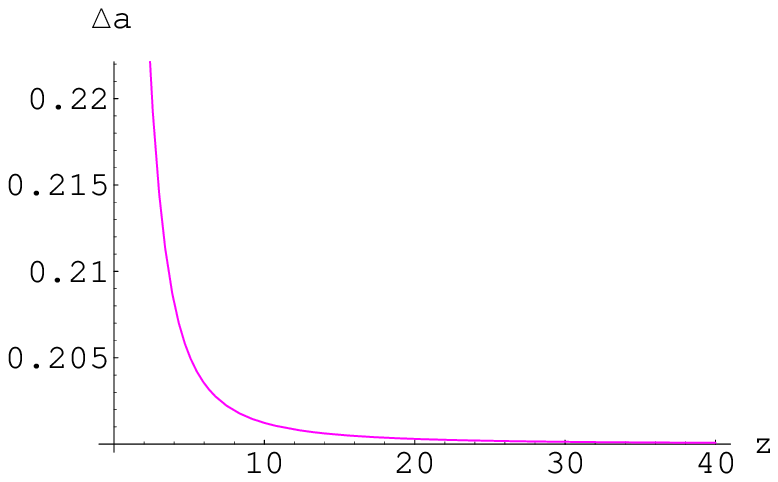,
width=0.45\linewidth}\caption{This graph is plotted for $h_2=1,
z_0=1$.}
\end{figure}
\begin{eqnarray}\nonumber
8\pi\mu
&=&\frac{4F}{(z^2+{z_0}^2)}+\frac{(z^2+{z_0}^2)^2}{F^2}+\frac{(z^2+{z_0}^2)({z_0}^2-3z^2)}{F^4}+\frac{8z^2(z^2+{z_0}^2)}{F^5},\\\label{c6}
\\\label{c8}
8\pi
P_{z}&=&\frac{4F}{(z^2+{z_0}^2)}+\frac{(z^2+{z_0}^2)^2}{F^2}+\frac{8z^2(z^2+{z_0}^2)}{F^5},
\\\label{c9}
8\pi
P_{\perp}&=&\frac{-6F}{(r^2+{z_0}^2)}+\frac{(r^2+{z_0}^2)({z_0}^2-3r^2)}{F^4}+\frac{12r^2(r^2+{z_0}^2)}{F^5}.
\end{eqnarray}
The dimensionless measure of anisotropy defined by Eq.(\ref{9}) in
this case takes the following form
\begin{equation}
\triangle
a=\frac{F^6-F(z_0^2-3{z}^2)(z^2+{z_0}^2)^2+\frac{z^2(3+8(z^2+{z_0}^2)^2)}{(z^2+{z_0}^2)}}{\left(4F^6+8z^2(z^2+{z_0}^2+F^3(z^2+{z_0}^2)^3)\right)}.
\end{equation}
All these quantities are shown graphically in figures \textbf{3} and
\textbf{4}.

\section{Junction Conditions}

In this section, we formulate junction conditions for the general
spacetime in the interior and vacuum spacetime in the exterior
regions, respectively. The quantities in the interior and exterior
are denoted by $-$ and $+$, respectively. The line element of the
exterior region is (Chao-Guang1995)
\begin{equation}\label{2.4.1}
ds^{2}_{+}=-NdT^{2}+{N}^{-1}dZ^{2}+Z^{2}(dX^2+dY^2),
\end{equation}
where $N(Z)=(\frac{-2M}{Z^{2}})$.

Now using the first junction condition, the continuity of the
intrinsic curvature (the first fundamental form)
\begin{equation}\label{2.4.6}
(ds^{2}_{-})_{\Sigma}=(ds^{2}_{+})_{\Sigma},
\end{equation}
we have the following equations
\begin{eqnarray}\label{M1}
\frac{dt}{d\tau}&=&\frac{1}{A}, Z_{\Sigma}=B,
\\\label{M2}
\frac{dT}{d\tau}&=&(\frac{-2M}{Z_{\Sigma}})^{\frac{1}{2}}
[(\frac{-2M}{Z_{\Sigma}})^{2}-
(\frac{dZ_{\Sigma}}{dT})^{2}]^{\frac{-1}{2}},
\end{eqnarray}
where $\tau$ is proper time defined for the boundary surface.

The second junction condition is the continuity of the extrinsic
curvature (the second fundamental form). The surviving components of
the extrinsic curvature for the interior spacetime are found as
follows
\begin{eqnarray}\label{2.4.14}
K_{00}^{-}&=&-[\frac{A'}{AC}]_{\Sigma},\\\
K_{11}^{-}&=&[\frac{BB'}{C}]_{\Sigma}=K_{22}^{-}.
\end{eqnarray}

The non-null components of the extrinsic curvature for the exterior
spacetime can be found as follows
\begin{eqnarray}\label{2.4.17}
K_{00}^{+}&=&[\frac{dZ}{d\tau}\frac{d^{2}T}
{d\tau^{2}}-\frac{d^{2}Z}{d\tau^{2}}\frac{dT}{d\tau}-
\frac{N}{2}\frac{dN}{dZ}(\frac{dT}{d\tau})^{3}+
\frac{3}{2N}\frac{dN}{dZ}\frac{dT}{d\tau}(\frac{dZ}{d\tau})^{2}]_{\Sigma}.\\\
K_{11}^{+}&=&[ZN \frac{dT}{d\tau}]_{\Sigma}=K_{22}^{+}.
\end{eqnarray}
By the continuity of extrinsic curvature components, we have the
following two equations
\begin{equation}\label{2.4.19}
-[\frac{A'}{AC}]_{\Sigma}=[\frac{dZ}{d\tau}\frac{d^{2}T}
{d\tau^{2}}-\frac{d^{2}Z}{d\tau^{2}}\frac{dT}{d\tau}-
\frac{N}{2}\frac{dN}{dZ}(\frac{dT}{d\tau})^{3}+
\frac{3}{2N}\frac{dN}{dZ}\frac{dT}{d\tau}(\frac{dZ}{d\tau})^{2}]_{\Sigma},
\end{equation}
\begin{equation}\label{2.4.20}
[\frac{BB'}{C}]_{\Sigma}=[ZN \frac{dT}{d\tau}]_{\Sigma}.
\end{equation}
Solving the above equations with Eqs.(\ref{M1}) and (\ref{M2}), we
get
\begin{equation}\label{2.4.29}
M\overset{\Sigma}{=}m(t,z).
\end{equation}
This is the necessary and sufficient condition for the matching of
interior and exterior spacetimes.

\section{Conclusion}
In this paper, we have studied the generating solution of field
equations with anisotropic in plane symmetric geometry. These
solutions are only possible due to the null heat flux in the source.
The sectional curvature mass as well as trapped surfaces have been
studied in detailed. The condition ${B'}=B^{2\alpha}$, implies the
absence of horizons. Under this condition density becomes decreasing
function and fluid maintains its anisotropy, the expansion scalar
becomes $\Theta=(2+\alpha)R^{(1-\alpha)}$. This implies that
$\Theta=0$, for $\alpha=-2$, $\Theta>0$, for $\alpha\geq0$,
$\Theta<0$, for $\alpha<-2$, which corresponds to bouncing,
expansion and collapse, respectively, which corresponds to bouncing,
expansion and collapse, respectively. For bouncing all density
becomes constant while pressure components are non-zero and zero,
which implies that anisotropy remains in the fluid. For collapse the
value of $\alpha $ is $\alpha=\frac{-5}{2}$ which implies that
$\Theta<0$ and all the matter components and anisotropy are
decreasing function of $z$ as shown in figures \textbf{1} and
\textbf{2}. To discuss the expansion, we have taken the value
$\alpha=\frac{3}{2}$, which implies that $\Theta>0$ and all the
matter components and anisotropy are decreasing function of $z$ as
shown in figures \textbf{3} and \textbf{4}.

The interior source metric has been matched with the exterior vacuum
solution by using the Darmois (1927) junction conditions. The detail
analysis of these conditions implies the continuity of interior mass
and exterior gravitational mass over the boundary surface.

\end{document}